\documentclass[
aps,
prb,
floatfix,
longbibliography,
reprint,
superscriptaddress
]{revtex4-2}

\usepackage[T1]{fontenc}
\usepackage{amsmath}
\usepackage{amssymb}
\usepackage{amsfonts}
\usepackage{mathptmx}
\usepackage{graphicx}
\usepackage{color}
\usepackage[dvipsnames]{xcolor}
\usepackage{bbm}
\usepackage[breaklinks=true,colorlinks=true,linkcolor=blue,urlcolor=blue,citecolor=blue]{hyperref}
\usepackage{array}
\usepackage{booktabs}
\usepackage{placeins}

\DeclareMathAlphabet{\mathcal}{OMS}{cmsy}{m}{n}

\begin{document}

\title{Trapping magnetic flux quanta in superconducting 3D caps}

\author{Igor Bogush}
\email{igori.bogus@tu-braunschweig.de}
\affiliation{Cryogenic Quantum Electronics, Institute for Electrical Measurement Science and Fundamental Electrical Engineering, Technische Universit\"at Braunschweig, Hans-Sommer-Str. 66, D-38106 Braunschweig, Germany}
\affiliation{Laboratory for Emerging Nanometrology (LENA), Technische Universit\"at Braunschweig, Langer Kamp 6A-B, D-38106 Braunschweig,  Germany}

\author{Vladimir M. Fomin}
\affiliation{Institute for Emerging Electronic Technologies, Leibniz IFW Dresden, Helmholtzstraße 20, 01069 Dresden, Germany}
\affiliation{Faculty of Physics and Engineering, Moldova State University, str. Alexei Mateevici 60, MD-2009 Chișinău, Republic of Moldova}

\author{Oleksandr Dobrovolskiy}
\affiliation{Cryogenic Quantum Electronics, Institute for Electrical Measurement Science and Fundamental Electrical Engineering, Technische Universit\"at Braunschweig, Hans-Sommer-Str. 66, D-38106 Braunschweig, Germany}
\affiliation{Laboratory for Emerging Nanometrology (LENA), Technische Universit\"at Braunschweig, Langer Kamp 6A-B, D-38106 Braunschweig,  Germany}
\affiliation{FLUXONICS---The European Foundry for Superconducting Electronics e.V., 38106 Braunschweig, Germany}

\begin{abstract}
Magnetic flux quanta in type-II superconductors are topological excitations of the order parameter whose sensitivity to the local environment makes them probes of current density, magnetic fields, and pinning landscapes. Yet, in planar thin films, individual vortices are difficult to manipulate, often requiring intricate nanopatterning or demanding scanning-probe instrumentation. Here, we show that shaping a superconducting thin film into a three-dimensional curved geometry enables controllable vortex trapping under a spatially uniform applied magnetic field. Employing a conformal formulation of the time-dependent Ginzburg–Landau equation, we simulate vortex dynamics in a cap-shaped superconducting membrane and compare it with a planar reference. Curvature converts the uniform applied magnetic field into a nonuniform local normal component, creating a reconfigurable asymmetric rim-pinning potential tunable by magnetic-field orientation. Rotating the magnetic field therefore controls the number and positions of vortices trapped along the rim. Moreover, the asymmetric pinning produces a superconducting diode effect through current-direction-dependent vortex capture. Three-dimensional curvature thus provides a route to engineer local vortex pinning and realize fluxonic devices with magnetic-field-programmable vortex trapping and transport.
\end{abstract}

\maketitle

\section{Introduction}
Magnetic flux penetrates type-II superconductors in the form of quantized flux lines---Abrikosov vortices, or fluxons~\cite{Abr57jpc}---topological defects of the superconducting order parameter stabilized by a conserved topological charge. Their sensitivity to local electromagnetic and pinning environments makes vortices nanoscale probes: their nucleation, position, and motion encode the local current density, magnetic field, and pinning landscape. Topological robustness also makes vortices attractive information carriers for fluxonic devices~\cite{Gol15nac}. Harnessing these capabilities, however, requires going beyond vortex imaging~\cite{Aus09nph,Emb17nac} to achieve control over individual vortex placement.

Existing strategies for vortex control span two regimes. At the collective level, engineered pinning landscapes tailor the vortex response~\cite{vel08jmm,pal12prb,wor17psr,zhu04prl,luk06jap,mil05jlt,Kar24pra,Bez22prb}, while directional transport can be induced by asymmetric pinning ratchets~\cite{lee99nat,Vil03sci,des06prb,jin10prb,zhu03prb}, magnetic textures and structures~\cite{ver08apl,vla08prb,cie13prb,sil11sst,Sou07prl}, and current and field gradients~\cite{xue17jpc,ker23nal}. At the single-vortex level, scanning-probe techniques enable deterministic manipulation using local magnetic fields~\cite{gar02apl}, tip dragging~\cite{str08apl,rei07phc,gre22pra}, or mechanical stress~\cite{Kre16nal}, with optical tweezing~\cite{ves16nco} and thermal gradients~\cite{jin18sst} providing non-contact alternatives. These approaches, however, face a trade-off: single-vortex manipulation generally requires dedicated instrumentation or intricate nanostructuring, whereas collective approaches offer limited control over individual vortex number and position. Dynamically reshaping the pinning landscape would bridge this gap, combining local addressability with global control.

Departing from planar geometry offers such a strategy \cite{Fom22apl}. Three-dimensional (3D) curved superconducting nanoarchitectures can reshape vortex matter through geometry alone \cite{Rez20cph,dee25acs}, without relying on engineered pinning sites or artificial defects \cite{vel08jmm,pal12prb,wor17psr,zhu04prl,luk06jap,mil05jlt,Kar24pra,Bez22prb}. In a homogeneous, moderate magnetic field, a thin superconducting film responds primarily to the local field component normal to its surface, $B_n=\mathbf{B}\cdot\hat{n}$, where $\hat{n}$ is the outward-pointing local surface-normal unit vector~\cite{Fom12nal}. For a planar film, $B_n$ is spatially uniform, whereas curvature converts the same homogeneous applied field into a position-dependent normal component~\cite{Fom21boo}. For instance, it has been demonstrated that magnetic field can manipulate curved phase-slip events in bowl-shaped 3D superconducting nanoarchitectures~\cite{BF2021} and control the directionality of fluxon transport in 3D superconducting open nanotubes~\cite{Bog24nan}. The local surface orientation therefore becomes an additional control parameter that is absent in planar geometries. Such 3D superconducting nanoarchitectures can be fabricated by focused electron- and ion-beam-induced deposition (FEBID/FIBID)~\cite{cor19nan,Por23acs,Hof23arx,Fer20mat,Cor18nal}, complementing strain-engineered self-rolling approaches for tubular and helical structures~\cite{Sch01nat,Fom21boo,Thurmer10}. Experimentally realized 3D superconductors already exhibit enhanced microwave response~\cite{Loe19acs}, discrete resistance transitions~\cite{cor19nan}, and non-local magnetoresistance~\cite{zha26afm}. These observations demonstrate that curvature is not merely a geometric modification, but a practical means of engineering the superconducting response.

Here, we apply a conformal formulation of the time-dependent Ginzburg–Landau (TDGL) model~\cite{Bog25cpc} to a Nb–C cap-shaped membrane, using a flat membrane as a reference.  The curved geometry converts an applied in-plane magnetic field component into a direction-dependent vortex-pinning potential at the cap rim, enabling controlled trapping of a well-defined number of vortices over a current window substantially wider than that for a purely out-of-plane field or a flat membrane. The resulting asymmetric pinning produces polarity-dependent vortex trapping and a superconducting diode effect~\cite{Gut23nac}. Finally, we show that trapped vortices can be steered around the entire cap circumference simply by rotating the in-plane magnetic field, even at zero transport current. These results establish curved 3D geometry as a design parameter for engineering local vortex-pinning landscapes and screening-current distributions in superconducting thin-film membranes, opening a route toward fluxonic devices with field-programmable vortex trapping and transport.

\section{Results}
\label{sec:results}

\subsection{Model}

Two device geometries are compared throughout this work: a cap-shaped membrane [Fig.~\ref{fig:1}, bottom] and a flat reference membrane [Fig.~\ref{fig:1}, top].
At the center of the cap-shaped membrane, a rotationally symmetric ``crater'' of 300\,nm depth extends below the surrounding flat plane [see the inset of Fig.~\ref{fig:1}].
Each device is contacted by two metallic leads carrying a transport current $I$ and is placed in an external magnetic field $\mathbf{B}$ applied in the $yz$ plane at an angle $\alpha$ with respect to the out-of-plane $z$ axis.
Both membranes are $1\,\mu$m wide and $50\,\mathrm{nm}$ thick and are modeled using the material parameters of FIBID-grown Nb-C strips~\cite{Por19acs}, summarized in Table~\ref{tab:params}.

\begin{table}[h!]
\caption{Material parameters of Nb-C \cite{Por19acs} used in the simulations.}
\label{tab:params}
\begin{ruledtabular}
\begin{tabular}{lcc}
Parameter & Expression & Value \\
\hline
Critical temperature & $T_\mathrm{c}$ & 5.6~K \\
Coherence length & $\xi(0)$ & 6~nm \\
Penetration depth & $\lambda(0)$ & 1~$\mu$m \\
GL parameter & $\kappa = \lambda/\xi$ & 167 \\
Normal-state resistivity & $\rho_n$ & 500~$\mu\Omega\cdot$cm \\
Relaxation time at 5 K & $\tau=\pi\hbar/[8k_\mathrm{B}(T_\mathrm{c}-T)]$ & 5.1~ps \\
Coherence length at 5 K & $\xi(T) = \frac{\xi(0)}{\sqrt{1-T/T_c}}$ & 18.3~nm \\
Penetration depth at 5 K & $\lambda(T) = \frac{\lambda(0)}{\sqrt{1-T/T_c}}$ & 3~$\mu$m \\
Pearl length at 5 K & $\Lambda(T) = 2\lambda^2(T)/d$ & 373~$\mu$m \\
Critical current at 5 K\footnotemark[1] & $I_\mathrm{c}(T)=\frac{\hbar wd}{3\sqrt3 e\mu_0\lambda^2(T)\xi(T)}$ & 29.5~$\mu$A\\
Upper critical field at 5 K & $B_\mathrm{c2}(T)=\Phi_0/\left(2\pi\xi^2(T)\right)$ & 1~T\\
\end{tabular}
\end{ruledtabular}
\footnotetext[1]{For a membrane cross-section of $wd=1000\times50$~nm$^2$.}
\end{table}

To analyze the spatiotemporal evolution of the superconducting order parameter, we numerically solve the TDGL equation using a finite-difference method combined with the conformal approach~\cite{Bog25cpc}. The calculations are performed within the thin-film and isothermal approximations, with magnetic stray fields neglected. The procedure, its justification and limitations are detailed in Appendix~\ref{sec:methods}. We next demonstrate several effects that emerge uniquely from the 3D cap geometry.

\subsection{Vortex trapping}
\begin{figure}
\centering
\includegraphics[width=1\columnwidth]{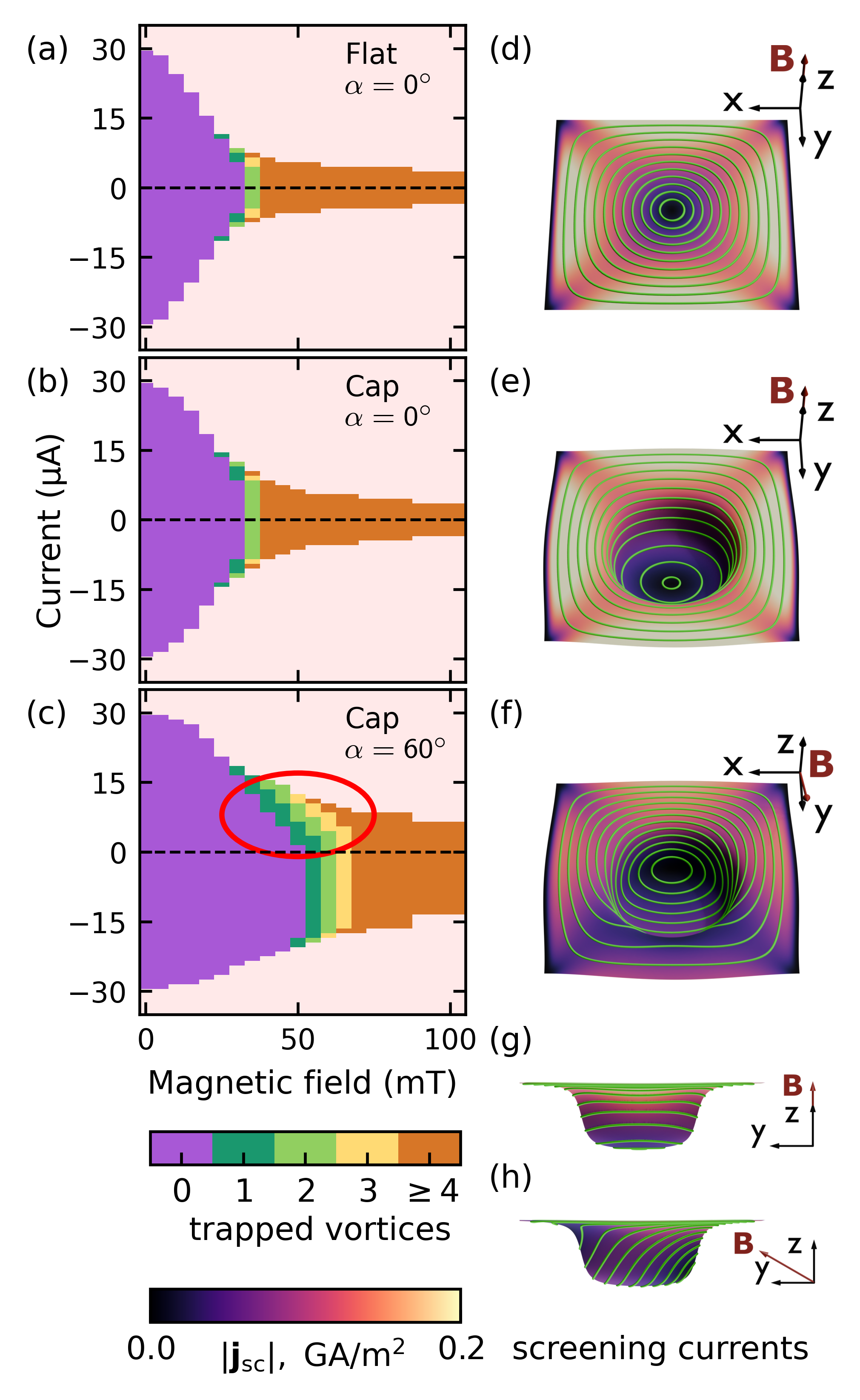}
\caption{
(a–c) Number of pinned vortices, $n_\mathrm{p}$, in the membrane as a function of magnetic field $B$ and transport current $I$: (a) flat membrane, $\alpha = 0^\circ$; (b) cap-shaped membrane, $\alpha = 0^\circ$; (c) cap-shaped membrane, with $\mathbf{B}$ tilted by $\alpha = 60^\circ$ from $z$. The red ellipse in (c) indicates the region of effective vortex trapping, exhibiting pronounced non-reciprocity between the two current polarities. The light-red region denotes the regime in which vortices are mobile and generate a finite voltage, or the system is switched to the normal state. (d--h) Screening-current streamlines induced by an external magnetic field $B = 10$\,mT for the geometries and field orientations corresponding to (a--c). Color indicates the magnitude of the screening-current density, $|\mathbf{j}_{\mathrm{sc}}|$.
}
\label{fig:2}
\end{figure}

Figure~\ref{fig:2} illustrates how curvature and the resulting screening-current distribution enable vortex trapping in the cap-shaped membrane. For a vortex to become trapped in the membrane, it must first nucleate and subsequently reach a stable equilibrium position, where it remains without denucleating or annihilating. If vortices keep nucleating without being trapped, they instead move in a flux-flow regime, resulting in a resistive state. At low magnetic fields, the screening currents are insufficient to lower the effective barrier for vortex nucleation \cite{Plo01prb}, requiring an additional bias from the transport current. At higher fields, the screening currents alone sufficiently reduce this barrier \cite{Sil25apr}, enabling vortex nucleation even at zero transport current. However, establishing a regime in which a nucleated vortex remains stably trapped, without further motion or denucleation, is more subtle \cite{Kre16nal}. Achieving controlled trapping of a prescribed number of vortices through the transport-current bias is even more challenging \cite{Emb17nac}.

\begin{figure}[t]
\centering
\includegraphics[width=0.96\linewidth]{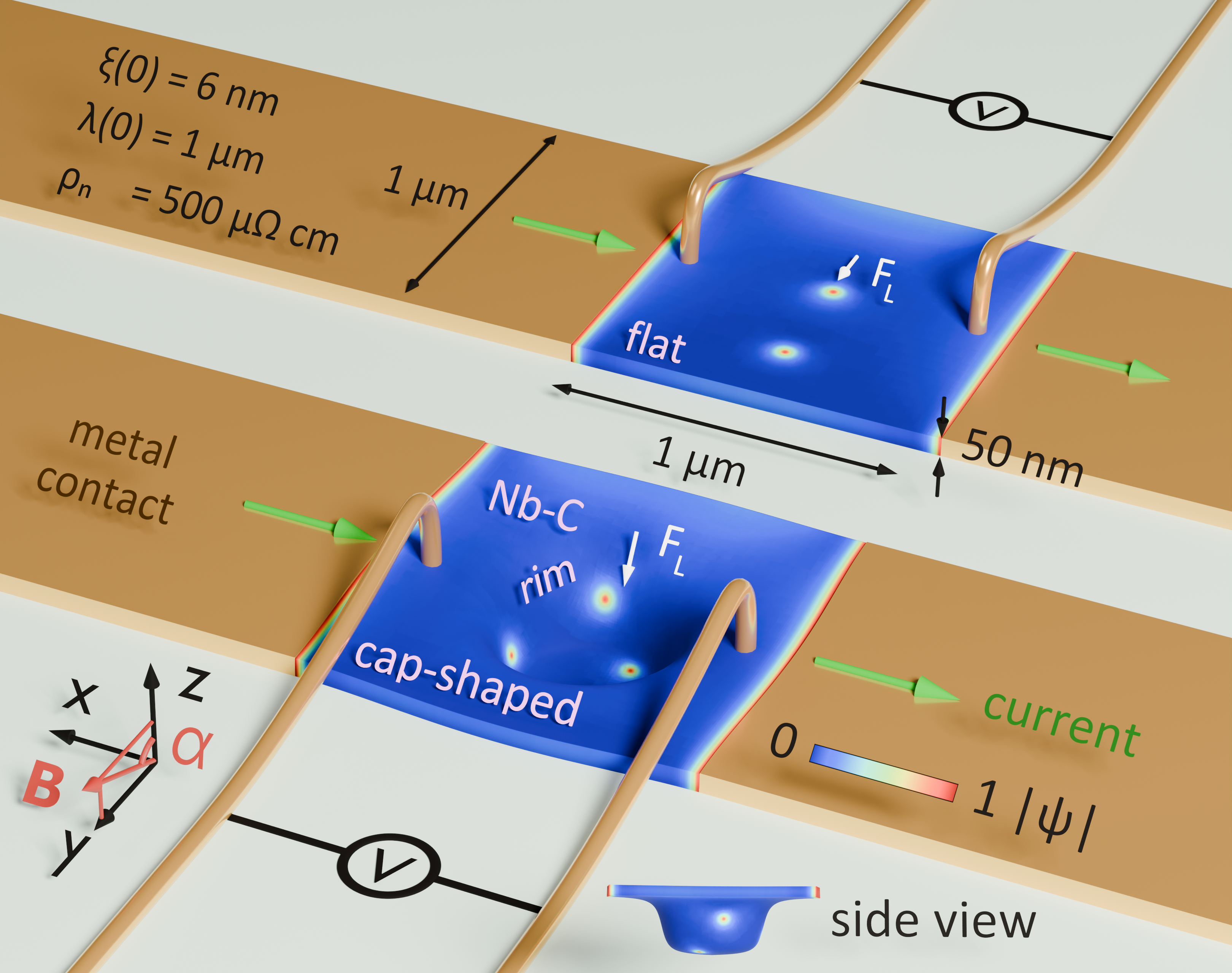}
\caption{
A cap-shaped membrane (bottom) and a flat reference membrane (top), each contacted by two current leads carrying a transport current $I$ and two voltage leads. A magnetic field $\mathbf{B}$ is applied in the $yz$ plane at an angle $\alpha = 60^\circ$ to the $z$ axis, with $|\mathbf{B}| = 50$\,mT and $I = 14\,\mu$A. Color maps show the magnitude of the superconducting order parameter, $|\psi|$. Inset: side view of the cap-shaped membrane.
}
\label{fig:1}
\end{figure}

In Fig.\,\ref{fig:2}, the number of vortices trapped in the superconducting membrane, $n_\mathrm{p}$, is mapped as a function of field magnitude $B$ and transport current $I$ for three cases: the flat membrane under an out-of-plane field [Fig.~\ref{fig:2}(a)], the cap-shaped membrane under an out-of-plane field [Fig.~\ref{fig:2}(b)], and the cap-shaped membrane under a field tilted by $\alpha=60^\circ$ from $z$ [Fig.~\ref{fig:2}(c)]. The tilt angle $\alpha=60^\circ$ was chosen as a representative value that provides a substantial in-plane field component while retaining a finite out-of-plane component, thereby making the curvature-induced asymmetry and vortex-trapping bands clearly visible. The effects discussed below are not specific to $\alpha=60^\circ$: finite field tilts produce the same qualitative mechanism, while the width and location of the trapping regions vary quantitatively with $\alpha$. No optimization of the tilt angle is implied; a systematic mapping of these boundaries lies beyond the scope of the present work.

Under the out-of-plane field ($\alpha = 0^\circ$), the flat and cap-shaped membranes [Fig.~\ref{fig:2}(a,b)] show essentially the same behavior: the map is symmetric under current reversal, and single-vortex trapping is achieved only within a narrow window of transport currents. Thus, in neither geometry does the out-of-plane field afford practical control over the trapped vortex number.

By contrast, the cap-shaped membrane in a field tilted by $\alpha=60^\circ$ [Fig.~\ref{fig:2}(c)] reveals a qualitative distinction between flat and curved geometries. Within the thin-film approximation, tilting the field for a flat membrane is equivalent to reducing its out-of-plane component, such that the resulting map is simply a contracted version of that in Fig.~\ref{fig:2}(a). For the curved membrane, tilting the field reshapes the spatial distribution of its local normal component over the entire surface. The most striking feature appears at positive transport current $I>0$ near $B=50$\,mT, highlighted by the red ellipse. Throughout the manuscript, $I>0$ is defined as transport current flowing along the $-x$ direction, as indicated by the green arrows in Fig.~\ref{fig:1}; current flowing along $+x$ corresponds to $I<0$. For $I>0$, the current-induced Lorentz-like force points along $+y$, whereas for $I<0$ it points along $-y$. Thus, instead of the narrow single-vortex trapping region found for the flat geometry, a broad and clearly separated band with $n_\mathrm{p}=1$, $2$, and $3$ emerges for the 3D cap, allowing the number of trapped vortices to be selected by the transport current. For $I<0$, the trapping band remain broad, yet the number of trapped vortices can no longer be selected by the transport current.

To elucidate the mechanism behind the region of enhanced control over trapped vortices by the transport current, we plot the screening-current streamlines for each case under a weak magnetic field of 10 mT and zero transport current [Fig.~\ref{fig:2}(d--f)]. The higher current density at the edge, arising from the screening response to the applied field, promotes vortex nucleation. When a transport current is switched on, it sums constructively with this edge screening current when the two are unidirectional, reinforcing vortex nucleation, and destructively when they flow in opposite directions, suppressing it. In both the flat and cap-shaped geometries under an out-of-plane field, the screening-current lines form concentric closed curves centered on the membrane. These lines are symmetric under spatial mirror reflection, which ensures reciprocity under current reversal. The Lorentz-like force produced by these screening currents pushes vortices toward the critical point at the center, where the screening current vanishes, indicated by dark regions inside the innermost line, Fig.~\ref{fig:2}(d--f). When the magnetic field is tilted, the center of the current-streamline pattern shifts toward the rim, so the streamlines become \emph{denser} at one edge and \emph{sparser} at the opposite edge. This asymmetry acts as an effective vortex-ratchet potential~\cite{lee99nat,Vil03sci,des06prb,Shk11prb}, giving rise to non-reciprocal vortex nucleation. Thus, the magnetic-field tilt angle controls the difference in current density between the two opposite edges, where vortices nucleate and denucleate, respectively. This suggests a route to promote vortex nucleation while suppressing denucleation, or vice versa for the opposite transport currents.

\subsection{Vortex steering}
\begin{figure}[t]
\centering
\includegraphics[width=0.95\columnwidth]{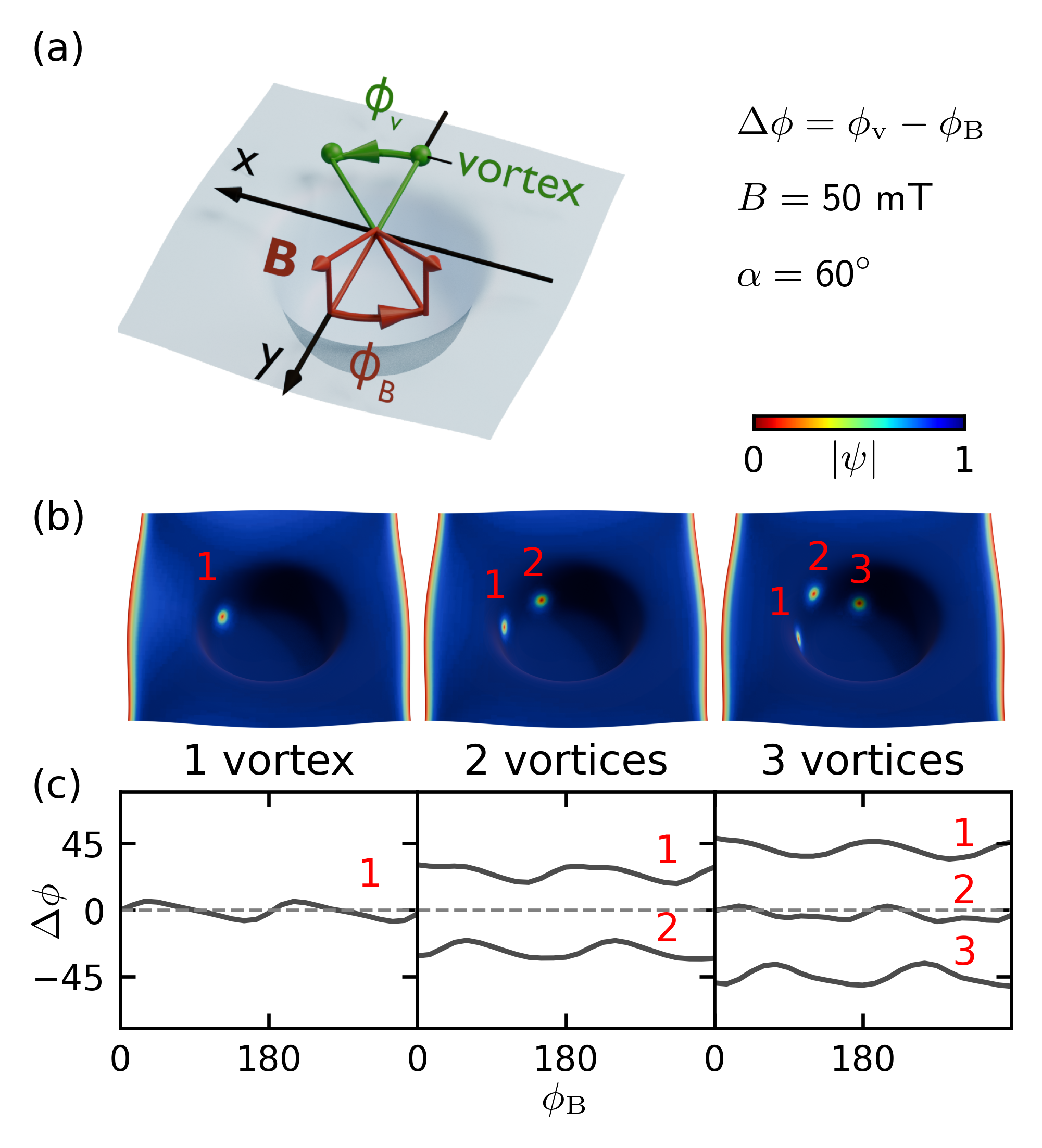}
\caption{(a) Schematic of vortex steering by rotation of the magnetic field in the $xy$ plane. The vortex azimuthal position $\phi_\mathrm{v}$ is measured with respect to the cap center, while $\phi_\mathrm{B}$ denotes the azimuthal angle of the magnetic-field direction.
(b) Order-parameter distribution $|\psi|$ for configurations with one, two, and three trapped vortices. The vortices are nucleated by dc-current pulses of 6, 9, and 12~$\mu$A, respectively, at a magnetic field of $B=50$~mT tilted by $\alpha=60^\circ$ with respect to the $z$ axis in the $yz$ plane. The current is subsequently reduced to zero, and the magnetic field is rotated in the $xy$ plane to $\phi_\mathrm{B}=45^\circ$.
(c) Angular offset $\Delta\phi=\phi_\mathrm{v}-\phi_\mathrm{B}$ as a function of $\phi_\mathrm{B}$ for the one-, two-, and three-vortex configurations.
}
\label{fig:3}
\end{figure}

Unlike vortices trapped at point-like or linearly extended pinning sites~\cite{Dob20pra,Kar24pra}, vortices confined by the 3D curved geometry remain steerable through the direction of the applied magnetic field. To demonstrate this, we consider a cap-shaped membrane with a $60^\circ$ field tilt, in which we first nucleate $n_\mathrm{p}=1$, $2$, and $3$ vortices at $B=50$~mT following the protocol of Fig.~\ref{fig:2}. We then set the transport current to zero, slowly rotate the magnetic field within the $xy$ plane by an azimuthal angle $\phi_{B}$ in steps of $15^\circ$, and track the azimuthal position $\phi_\mathrm{v}$ of each trapped vortex, as shown in Fig.~\ref{fig:3}(a). Notably, all vortices trapped in the cap remain pinned once the transport current is switched off, confirming that the trapping observed in Fig.~\ref{fig:2} is persistent rather than merely transient.

In an idealized rotationally symmetric membrane, vortices would follow the magnetic field with a constant angular shift $\Delta\phi = \phi_\mathrm{v} - \phi_\mathrm{B}$ between the field direction and the vortex position. Although the cap itself is rotationally symmetric, the square confinement geometry around the cap breaks this symmetry, producing the edge--vortex interactions that prevent $\Delta\phi$ from remaining strictly constant [Fig.~\ref{fig:3}(c)]. Despite this effect, in the single-vortex configuration the vortex closely tracks the field direction, deviating by no more than $|\Delta\phi| < 8^\circ$ [Fig.~\ref{fig:3}(c)]. In the two- and three-vortex configurations, the vortices exhibit an approximately constant angular shift arising from the vortex--vortex repulsion, deviating by no more than $9^\circ$ from their average values [Fig.~\ref{fig:3}(b,c)]. A full $360^\circ$ revolution of the field causes vortex trajectories to trace out complete circles, offering a field-driven alternative to the closed-loop vortex trajectories previously achieved via transport current in the Corbino geometry~\cite{lop99prl}. Taken together, these results show that vortices trapped due to the 3D curved geometry are not merely pinned but remain dynamically responsive to the magnetic-field direction: once nucleated, they stay trapped without transport current while being continuously steered around the cap rim.

\begin{figure}[t]
\centering
\includegraphics[width=1\columnwidth]{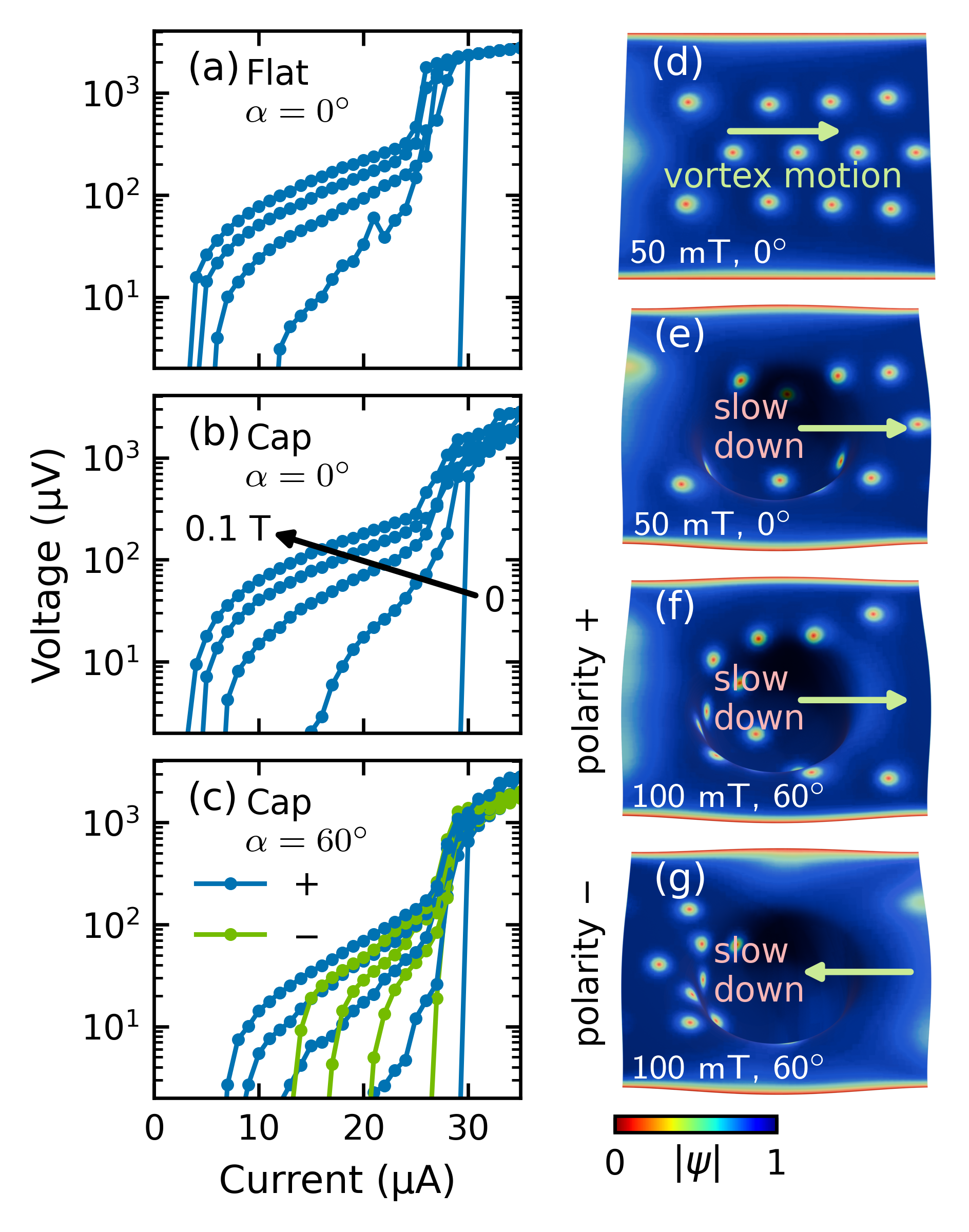}
\caption{
(a–c) $I$-$V$ curves at magnetic fields $B = 0, 25, 50, 75, 100$ mT for: (a) flat membrane, $\alpha = 0^\circ$; (b) cap-shaped membrane, $\alpha = 0^\circ$; (c) cap-shaped membrane, $\mathbf{B}$ tilted by $\alpha = 60^\circ$ from $z$. For the tilted field, the cap-shaped membrane exhibits non-reciprocity with respect to current reversal. (d–g) Order parameter distribution showing vortices at the indicated magnetic field and tilt angle, at $I=15$ $\mu$A. For the tilted magnetic field, its magnitude is increased so as to keep the $B_\mathrm{z}$ component fixed. Light green arrows indicate the direction of vortex motion; vortices slow down inside the cap-shaped region.
}
\label{fig:4}
\end{figure}

\subsection{Vortex-flow regime}
Trapped, immobile vortices generate no voltage, even under an applied transport current. Only once the transport current is increased further do vortices depin and enter a flux-flow regime, generating a measurable voltage that contains additional information about vortex dynamics.

Figure~\ref{fig:4}(a--c) shows the current-voltage ($I$-$V$) curves for the three cases discussed above, together with the corresponding order-parameter distributions at $I=15$~$\mu$A [Fig.~\ref{fig:4}(d--g)]. At first glance, the $I$-$V$ curves and trajectories for the flat and cap-shaped membranes under the out-of-plane field appear qualitatively similar. Nevertheless, vortices moving through the cap-shaped crater in the middle of the membrane slow down [Fig.~\ref{fig:4}(e--g)], since the current density there is reduced by the dimensional de-crowding effect~\cite{Bog26arx}. This results in a modest voltage reduction, from 50 to 37\,$\mu$V at 15\,$\mu$A. The de-crowding effect originates from the locally increased cross-section of the membrane associated with the cap extending below the surrounding flat plane. For the cap-shaped membrane under the tilted field, vortices inside the crater slow down further, or are even captured in the rim [Fig.~\ref{fig:4}(f,g)]. Additionally, non-reciprocity manifests in the $I$-$V$ curves [Fig.~\ref{fig:4}(c)], indicating a superconducting diode effect~\cite{Gut23nac}. For instance, using a $2\,\mu$V voltage criterion for the $I$-$V$ curves at $B=50$\,mT in Fig.~\ref{fig:4}(c), we obtain critical currents of $I_\mathrm{c}^{+}=12.5\,\mu$A and $I_\mathrm{c}^{-}=20.5\,\mu$A. This results in a diode efficiency of $\eta=\left||I_\mathrm{c}^{+}|-|I_\mathrm{c}^{-}|\right|/(|I_\mathrm{c}^{+}|+|I_\mathrm{c}^{-}|)\approx24\%$, which is approximately a factor of two smaller than the experimental efficiencies reported for optimized superconducting diode structures~\cite{And20nat,Hou23prl,mol25nat}. Accordingly, the value deduced for our 3D cap-shaped membrane should be regarded as representative rather than optimized, since the extracted critical currents and hence $\eta$ depend quantitatively on the chosen voltage criterion and magnetic field. Nevertheless, it falls within the experimentally established range despite the absence of engineered asymmetric pinning sites. The observed non-reciprocity is therefore an intrinsic consequence of the \emph{tilted-field response} of the cap-shaped membrane, manifesting both in its transport characteristics and in the individual vortex trajectories. 

\subsection{Prototype fluxon register}

\begin{figure}[t!]
\centering
\includegraphics[width=0.86\columnwidth]{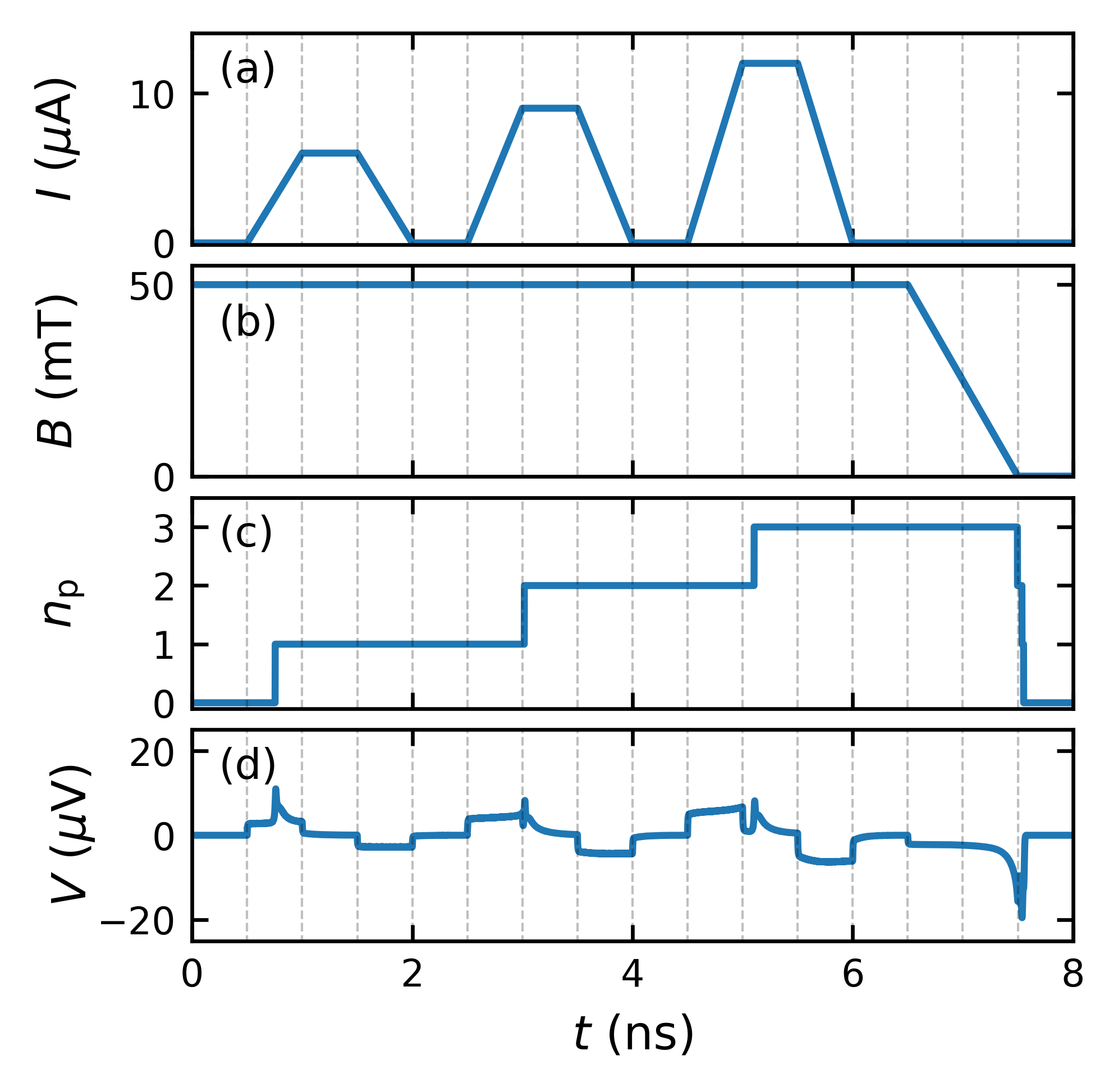}
\caption{Concept of a fluxon register based on a 3D superconducting cap, with the stored state encoded by the number of trapped vortices, $n_\mathrm{p}$. (a) Transport-current pulse sequence $I(t)$ used to sequentially inject vortices into the cap. (b) Magnetic-field magnitude $B(t)$, including the field down-ramp used to release the trapped vortices and reset the register. (c) Number of trapped vortices $n_\mathrm{p}(t)$, demonstrating stepwise writing and retention of the states $n_\mathrm{p}=0,1,2,3$. (d) Corresponding voltage response $V(t)$ across the membrane, providing an electrical signature of vortex entry and release.}
\label{fig:5}
\end{figure}

The effects discussed above open a route toward 3D fluxonic devices based on controlled vortex capture, storage, and release. As a proof-of-principle implementation, Fig.~\ref{fig:5} illustrates the concept of a fluxon register. The cap-shaped membrane acts as a geometrically defined vortex trap whose state is encoded by the discrete number of trapped flux quanta, $n_\mathrm{p}=0,1,2,3,\ldots$. Current pulses [Fig.~\ref{fig:5}(a)] sequentially inject vortices into the cap, allowing the stored state to be programmed, as reflected by the stepwise evolution of $n_\mathrm{p}$ in Fig.~\ref{fig:5}(c). The accompanying voltage response [Fig.~\ref{fig:5}(d)] provides an electrical signature of the vortex-entry events. Between the pulses, the vortices remain trapped after the transport current is returned to zero, providing retention of the programmed state. Finally, reducing the magnetic field [Fig.~\ref{fig:5}(b)] releases the trapped vortices and resets the device to the vortex-free state.

The 3D fluxon register thus combines multilevel flux storage with current-controlled writing, zero-current retention, and field-controlled erasure in a single geometrically engineered superconducting element, suggesting a route toward fluxonic memory~\cite{Gol23nac} and logic architectures~\cite{Gol15nac}. We emphasize that this constitutes a proof-of-principle demonstration within a deterministic TDGL model. The trapped states remain stable over the simulated time window, whereas the retention and robustness of a real device will additionally depend on thermal fluctuations, defects, inhomogeneities, and tolerance to field and current noise, which are beyond the scope of the present study. Additionally, the trapped magnetic flux could also be detected locally, for example by scanning SQUID microscopy~\cite{Kre16nal,Emb15scr}.

\section{Conclusion}
\label{sec:conclusion}

We have used conformal time-dependent Ginzburg-Landau simulations~\cite{Bog25cpc} to study the geometric trapping and steering of single and few vortices in a 3D cap-shaped superconducting thin-film membrane. We have shown that the cap-shaped membrane responds to the direction of an in-plane magnetic field component, developing an asymmetric effective pinning potential localized at the cap rim as a result of the curved geometry. This asymmetry enables the trapping of a controllable number of vortices over a wide current window, with clearly separated bands resolved for one, two, and three vortices. Trapped vortices do not leak out even after the transport current is set back to zero. This effective pinning, arising from the interplay between geometry and magnetic field, further allows trapped vortices to be steered continuously around the cap rim by rotating the in-plane field direction alone, at zero transport current. Once a vortex depins from the cap rim under a sufficiently strong transport current, a flux-flow regime is established, generating dissipation. The resulting voltage response is non-reciprocal under current reversal, since moving vortices traverse the asymmetric effective pinning potential at the rim.

The predictions of our studies are testable for Nb-C cap-shaped structures, which can be fabricated by direct-write FEBID/FIBID~\cite{cor19nan,Por23acs,Hof23arx}. The non-reciprocal transport with respect to current reversal can be measured via current-voltage characteristics, and the vortex configurations can be verified by direct imaging~\cite{Aus09nph,Emb17nac}. Real structures inevitably contain surface and edge defects, pinning centers, and other imperfections. However, the fine granularity of FEBID/FIBID materials is expected to average out strong local pinning~\cite{ser16bjn}, making the effects predicted here experimentally accessible.

In summary, the discussed results establish curved 3D geometry as a design parameter that tunably reshapes the effective pinning landscape and screening-current distribution of a superconducting film. This geometric control over individual vortices suggests a route toward 3D superconducting nanoarchitectures with potential applications for vortex-based memory \cite{Gol23nac} and logic applications \cite{Gol15nac}.

\acknowledgments
I.B. acknowledges support from the Deutsche Forschungsgemeinschaft (DFG, German Research Foundation) through Grant No. 583926508 (SyMPaThY) and the use of the CryoMind and CryoFryer simulation workstations at CryoQuant/TU Braunschweig. V.M.F. acknowledges support from the DFG through Grant No. 440761189. O.D. acknowledges support from the DFG through Grant No. 577881064 (Super3DMag). This research is based on work from the COST Actions CA16218 (NANOCOHYBRI), CA21144 (SuperQuMap) and CA23134 (Polytopo), supported by the European Cooperation in Science and Technology.

\section*{DATA AVAILABILITY STATEMENT}
The data supporting the findings of this study are openly available\,\cite{Bog26DSTrapping}.

\appendix
\section{Methods}
\label{sec:methods}

\subsection{Geometry and physical model}

We consider a thin-film superconducting membrane contacted by two normal-metal leads carrying a transport current $I$ and subjected to a spatially homogeneous external magnetic field $\mathbf{B}$ applied in the $yz$ plane at an angle $\alpha$ with respect to the $z$ axis (see Fig.~\ref{fig:1}). We compare two geometries: a flat square reference membrane with a side length of $1~\mu$m and a cap-shaped membrane containing a rotationally symmetric crater centered in the film plane, parametrized as
\begin{equation}
    z(x,y)=\frac{A}{\pi}\arctan\!\left(\frac{r-C}{B}\right),
    \qquad
    r=\sqrt{x^2+y^2},
    \label{eq:cap-surface}
\end{equation}
where $C=250$~nm is the rim radius, $B=25$~nm the transition width, and $A=300$~nm the cap depth. The geometry is defined over the square domain $x,y\in[-500,500]$~nm.

Both structures are modeled using parameters representative of FIBID-deposited Nb-C films \cite{Por19acs} with thickness $d=50$~nm. The validity of the thin-film approximation rests on three conditions. First, the order parameter must remain approximately uniform across the film thickness. Since it recovers over a distance of order $2\xi$ from a vortex core, this requires $d\lesssim4\xi(T)$; in 3D geometries, vortices may briefly become tangent to the membrane surface after nucleation, while the thin-film description remains applicable~\cite{Bog26arx}. Second, for a curved membrane, the effective 2D differential operators require $d<R_\pm$, which is satisfied here because the minimum radius of curvature of the cap is $\approx80$~nm. Third, the applied fields obey $B\ll B_\mathrm{c2}(T)=\Phi_0/[2\pi\xi^2(T)]$, where $\Phi_0$ is the flux quantum. For the present simulations at $T=5$~K, $B_\mathrm{c2}(5~\mathrm{K})$ is of the order of $1$~T for FIBID-deposited Nb-C \cite{Por19acs}. In this regime, the superconducting response is governed predominantly by the magnetic-field component normal to the membrane surface~\cite{Gen66boo}. This description is supported experimentally: in $100$~nm Nb films ($T_\mathrm{c}=8.7$~K, $\xi(0)\approx19$~nm), an in-plane field of $\sim100$~mT produced no noticeable change in the normal-to-surface flux dynamics at temperatures from $3.6$ to $7$~K ($d/\xi(3.6\,\text{K})\approx 4$), which the authors attribute to the absence of in-plane vortices in such thin films~\cite{vla16prb}.

We further neglect the magnetic self-field generated by the transport current, justified by the extreme type-II character of Nb-C ($\kappa\approx167$) and its large Pearl length~\cite{Pea66jap}, $\Lambda(5\,\mathrm{K})=2\lambda^2(5\,\mathrm{K})/d\approx373~\mu\mathrm{m}$, which exceeds the membrane's lateral dimensions by more than three orders of magnitude and implies nearly complete field penetration. Consistently, for the maximum current considered, $I=30~\mu$A, the self-field estimate for the planar membrane $B_{\mathrm{self}}=(\mu_0 I/2w)\ln(2w/d)\approx0.07$~mT for $w=1~\mu$m and $d=50$~nm is negligible compared with the applied magnetic fields used in the simulations.

We adopt the isothermal approximation throughout because heating during vortex nucleation and early flux flow is weak: at low currents, vortices move slowly, whereas appreciable self-heating arises only near the transition to the normal state, where rapid vortex motion dissipates substantial power \cite{Dob20nac}. For a substrate thermal boundary conductance $h\approx10^4$~W/m$^2$K at 5~K~\cite{Bez19prb}, the dissipated power remains below 5~nW throughout the first half of the flux-flow branches in Fig.~\ref{fig:4}, corresponding to a temperature rise of less than $0.5$~K. At higher currents, experiments may enter a highly resistive state earlier than predicted by the isothermal model because of self-heating \cite{Bez19prb,Dob20nac}. The used approximation is therefore appropriate for the present focus on low-current vortex control, where overheating is not dominant.

Finally, our model describes an idealized structure, neglecting edge roughness, inhomogeneity, and thermal fluctuations. FIBID-deposited membranes are in the dirty limit, where the vortex core is comparable to or larger than the typical crystallite size \cite{Dob20nac}, so no single crystallite acts as a strong, isolated pinning center. These imperfections may shift quantitative critical-current values and nucleation details, but since the pinning and steering mechanisms identified here depend only on geometry, they are expected to persist in real samples \cite{ser16bjn}. A systematic quantitative analysis of the effect of disorder is beyond the scope of the present work.

\subsection{TDGL equation}

Within the conformal approach~\cite{Bog25cpc}, we use a two-dimensional TDGL equation for the complex order parameter $\psi$,
\begin{equation}
    \tau \left(
        \partial_t + i \frac{2 e}{\hbar} \varphi
    \right) \psi
    =
      \xi^2 \left(
        \nabla - i \frac{2 e}{\hbar} \mathbf{A}
    \right)^2 \psi
    + (1 - |\psi|^2) \psi
\label{eq:tdgl}
\end{equation}
coupled to a Poisson equation for the electric scalar potential $\varphi$,
\begin{align}
\Delta\varphi = \sigma_n^{-1}\nabla\cdot\mathbf{j}_\mathrm{sc}, \label{eq:poisson}\\
\mathbf{j}_\mathrm{sc}
     = \frac{\hbar}{2 \mu_0 e \lambda^2}
     \Im\left\{
        \psi^*
        \left(
        \nabla - i \frac{2e}{\hbar}\mathbf{A}
        \right)\psi
    \right\},
\label{eq:jsc}
\end{align}
where $\mathbf{A}$ is the vector potential tangential to the surface, and $\sigma_\mathrm{n}$ the normal-state conductivity. Quasi-static electric and magnetic fields are $\mathbf{E} = - \nabla \varphi$ and $\mathbf{B} = \nabla\times\mathbf{A}$ respectively.

At the free (insulating) edges $\Gamma_\mathrm{f}$ of the simulated domain and at the current-carrying contacts $\Gamma_\mathrm{c}$,
\begin{align}
\mathbf{n} \cdot \left(\nabla-i\frac{2e}{\hbar}\mathbf{A}\right)\psi = 0,\ \ \mathbf{n} \cdot \nabla \varphi = 0 &\quad\text{on }\Gamma_\mathrm{f}, \label{eq:bc-free}\\
\psi = 0,\ \ \mathbf{n}\cdot\nabla\varphi = -j_\mathrm{tr}/\sigma_\mathrm{n} &\quad\text{on }\Gamma_\mathrm{c}, \label{eq:bc-contact}
\end{align}
with $\mathbf{n}$ the outward normal and $j_\mathrm{tr} = I / w d$ the injected current density.

\subsection{Numerical implementation}

Equations~(\ref{eq:tdgl})--(\ref{eq:bc-contact}) are discretized on the resulting rectangular grid using the finite-difference method, with gauge-invariant link variables for the covariant derivatives of $\psi$~\cite{Kat93prb}. The order parameter $\psi$ is advanced with an explicit time-stepping scheme, while the Poisson equation for $\varphi$ is solved iteratively at every time step using the bi-conjugate gradient stabilized method. The computational grid has $384\times384$ points, and the explicit scheme uses a time step $\Delta t=0.01$~ps.

Grid-point positions were optimized to form a mesh in isothermal coordinates using the \texttt{IsoCoord} library~\cite{Bog25cpc,repo}, after which the domain boundary was no longer exactly square, but insignificantly convex. Isothermal coordinates are orthogonal, with a constant ratio between the lengths of the two basis vectors; in this parametrization, differential operators take the same form as in Cartesian coordinates, up to an overall multiplicative conformal factor $\mathrm{e}^{2\sigma}$, which can be absorbed locally by a constant rescaling. Equations~(\ref{eq:tdgl})--(\ref{eq:bc-contact}) can therefore be treated, at each grid point, as flat equations written in an orthonormal basis tangent to the surface.

\subsection{Simulation protocol}

The simulations were performed as follows. The order parameter is initialized once as $\psi=1$ and $\varphi=0$ everywhere, with $\psi$ smoothed to zero at the current contacts. The system is then allowed to relax for 10~ps at zero field and zero current, after which the field $B$ is ramped to its target value over 90~ps, and the system is allowed to relax for a further 200~ps.
Within this same run, the transport current $I$ is subsequently ramped upward continuously from zero, in steps of $1$~$\mu$A, each step ramped over 40~ps and followed by a 960~ps relaxation interval and a further 1~ns wait, over which the voltage $U$ and the number of vortices pinned in the central region of the membrane (at the cap rim for the cap-shaped structure), $n_\mathrm{p}$, are extracted for that current step before the ramp proceeds to the next one. These results are used to construct the $I$-$V$ curves in Fig.~\ref{fig:4} and the vortex-number maps, $n_\mathrm{p}$, in Fig.~\ref{fig:2}. 
The voltage $U$ is computed as the difference of the scalar potential $\varphi$ between two probes located 100~nm from the midpoint of each current contact, rather than at the metal--superconductor interface itself. This voltage is averaged over the 1~ns wait period at each current step.

For the vortex-steering results in Fig.~\ref{fig:3}, a similar protocol is used. The order parameter $\psi$ is initialized as before and allowed to relax for 0.1~ns at zero field and zero current. The field $B$ is then ramped to its target value over 0.1~ns and the system is allowed to relax for a further 0.5~ns. The transport current $I$ is subsequently ramped up over 0.5~ns, held for a 1~ns relaxation interval, then ramped back down to zero over 0.5~ns and followed by a 1~ns relaxation period, nucleating and trapping the desired vortex configuration. Finally, the in-plane field direction is rotated within the $xy$ plane by $15^\circ$ over 0.1~ns, and the system is allowed to relax for 0.4~ns; this rotation-relaxation step is repeated until the field has completed a full $360^\circ$ rotation.

\bibliography{refs}

\end{document}